\documentclass[12pt,prb,onecolumn]{revtex4}
\usepackage{graphicx,amsmath,amssymb,times}
\begin{document}

\renewcommand{\Re}{\mathop{\mathrm{Re}}}
\renewcommand{\Im}{\mathop{\mathrm{Im}}}
\renewcommand{\b}[1]{\mathbf{#1}}
\renewcommand{\c}[1]{\mathcal{#1}}
\renewcommand{\u}{\uparrow}
\renewcommand{\d}{\downarrow}
\newcommand{\bsigma}{\boldsymbol{\sigma}}
\newcommand{\blambda}{\boldsymbol{\lambda}}
\newcommand{\tr}{\mathop{\mathrm{tr}}}
\newcommand{\sgn}{\mathop{\mathrm{sgn}}}
\newcommand{\sech}{\mathop{\mathrm{sech}}}
\newcommand{\diag}{\mathop{\mathrm{diag}}}
\newcommand{\half}{{\textstyle\frac{1}{2}}}
\newcommand{\sh}{{\textstyle{\frac{1}{2}}}}
\newcommand{\ish}{{\textstyle{\frac{i}{2}}}}
\newcommand{\thf}{{\textstyle{\frac{3}{2}}}}

\bibliographystyle{naturemag}

\title{Single valley Dirac fermions in zero-gap HgTe quantum wells}

\author{B. B\"{u}ttner$^{1}$, C.X. Liu$^{1}$, G. Tkachov$^{1}$, E.G. Novik$^1$, C. Br\"{u}ne$^{1}$, H. Buhmann$^{1}$, E.M. Hankiewicz$^{1}$, P. Recher$^{1}$, B. Trauzettel$^{1}$, S.C. Zhang$^{2}$ and L.W. Molenkamp$^{1}$}

\affiliation{$^1$Faculty for Physics and Astronomy and R\"{o}ntgen Center for Complex Material Systems, Universit\"at W\"urzburg, Am Hubland, D-97074, W\"urzburg, Germany\\
$^2$ Department of Physics, McCullough Building,
Stanford University, Stanford, CA 94305-4045}

\begin{abstract}
Dirac fermions have been studied intensively in condensed matter physics in recent years.
Many theoretical predictions critically depend on the number of valleys where the Dirac
fermions are realized. In this work, we report the discovery of a two dimensional
system with a single valley Dirac cone. We study the transport properties of HgTe quantum
wells grown at the critical thickness separating between the topologically trivial and the quantum spin Hall phases. At high magnetic fields, the quantized Hall plateaus demonstrate the presence of a single valley Dirac point in this system. In addition, we clearly observe the linear dispersion of the zero mode spin levels. Also the conductivity at the Dirac point and its temperature dependence can be understood from single valley Dirac fermion physics.
\end{abstract}

\maketitle
\section{Introduction}
In recent years, Dirac fermions have been intensively studied in a number of condensed matter
systems. In the two dimensional material graphene the low energy spectrum is well described
by two spin degenerate massless Dirac cones at two inequivalent valleys, giving rise to four
massless Dirac cones in total\cite{semenoff1984,divincenzo1984}. The fabrication of graphene sheets enabled
substantial experimental progress in this field, and the physics of the Dirac fermions has been
investigated extensively\cite{castro2009}.  At the same time, many theoretical predictions rely
on a single Dirac cone valley, or, at least, weak inter-valley scattering\cite{castro2009}.
Graphene is not a suitable platform to test these latter predictions because of the presence of
two valleys and strong inter-valley scattering. In addition, it is presently unclear how
an energy gap can be reliably generated in single layer graphene, which would be desirable for a variety of device applications.

A HgTe/CdTe quantum well is another system where Dirac fermion physics emerges\cite{bernevig2006d,koenig2007}. In this case, the Dirac fermions appear only at a single valley, at the $\Gamma$ point of the Brillouin zone. Furthermore, tuning the thickness $d$ of the HgTe quantum well continuously changes both the magnitude and the sign of the Dirac mass. When $d$ is less than a critical
thickness $d_c\simeq 6.3$ nm, the system is in a topologically trivial phase with a full
energy gap. On the other hand, when $d>d_c$, the quantum spin Hall state is realized, where a
full energy gap in the bulk occurs together with  gapless spin-polarized states at the edge. The experimental discovery of this state\cite{koenig2007} provides the first example of a time-reversal invariant topological insulator in nature\cite{koenig2008}. A topological quantum phase transition is predicted to occur when $d=d_c$, where a massless Dirac fermion state is realized at a single valley, with both spin orientations\cite{bernevig2006d}. Our paper reports the experimental discovery of such a state.

In a two dimensional system with time reversal symmetry and half integral spin, a minimal
number of two massless Dirac cones can be present, as can be proven by a simple
generalization of a similar theorem in one dimension\cite{wu2006}. General results of this type have first been discovered in lattice gauge theory, and are known as chiral fermion doubling theorems\cite{nielsen1981}. In this sense, the HgTe quantum well at the critical thickness $d=d_c$ realizes this minimal number of two Dirac cones in two dimensions\cite{footnote}.

\section{HgTe quantum wells as half-graphene}
\begin{figure}
   \centering
   \includegraphics[width=0.6 \textwidth]{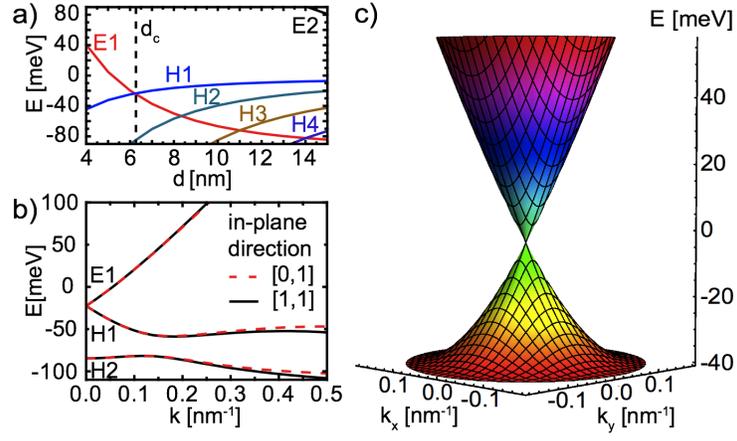}
    \caption{ (a) Sub-band energies of  HgTe/Hg$_{0.3}$Cd$_{0.7}$Te quantum wells as a function of well thickness $d$.
    (b) In plane dispersion of a quantum well at the critical thickness $d_c\simeq 6.3 $ nm. (c) A 3D plot of the Dirac cone describing the low energy spectrum for $d=d_c$.  }
    \label{fig:Dirac}
\end{figure}

HgTe is a zinc-blende-type semiconductor with an inverted band structure.
Unlike conventional zinc-blende semiconductors, and due to the very strong spin-orbit coupling
in the material, the $\Gamma_8$ band of HgTe (which derives from chalcogenide p-orbitals), has a higher energy than the $\Gamma_6$ band that originates from metallic s-orbitals and usually acts as the conduction band.
Consequently, in HgTe/(Hg,Cd)Te quantum wells, when the well thickness is large
enough, the sub-bands of the quantum well are also inverted:
$\Gamma_8$ derived heavy hole-like (H) sub-bands have higher energies
than $\Gamma_6$-based electron-like (E) sub-bands.
The inverted band structure, especially the
inversion between $E_1$ and $H_1$ sub-bands (where the suffix is the sub-band number index), leads to
the occurrence of the quantum spin Hall effect\cite{bernevig2006d,koenig2007}, boasting dissipationless
edge channel transport at zero external magnetic field\cite{roth2009}.
When the thickness of the quantum well is decreased, the energies of the $E$ sub-bands
increase due to quantum confinement, while those of the $H$ sub-bands decrease,
as shown in Fig \ref{fig:Dirac} (a). Eventually,
the $E_1$ sub-band gains a higher energy than the $H_1$ sub-band
and the system has the normal band sequence.
The different dependence of $E$ and $H$ sub-bands on well thickness
implies that there must exist a critical thickness where the band gap is closed.
In fact, the crossing point between $E_1$ and $H_1$ sub-bands,
denoted as $d_c$ in Fig \ref{fig:Dirac} (a), not only
corresponds to the critical point for the quantum phase transition
between quantum spin Hall insulator and normal insulator\cite{bernevig2006d}
but additionally yields a quantum well whose low-energy band structure closely
mimics a massless Dirac Hamiltonian.

When $d=d_c\simeq 6.3$ nm, the energy dispersion of the $E_1$ and $H_1$ sub-bands,
which can be calculated from an 8-band Kane model,
is found to linearly depend on the momentum $k$ near the $\Gamma$ point of the Brillouin zone, as
shown in Fig \ref{fig:Dirac} (b) and (c). Near the $\Gamma$ point,
using the states $|E_1,\frac{1}{2}\rangle$,
$|H_1,\frac{3}{2}\rangle$, $|E_1,-\frac{1}{2}\rangle$ and $|H_1,-\frac{3}{2}\rangle$ as a basis,
one can write an effective Hamiltonian for the $E_1$ and $H_1$ sub-bands, as follows\cite{bernevig2006d}
\begin{eqnarray}
	&&H_{eff}(k_x,k_y)=\left(
	\begin{array}{cc}
		H_D({\bf k})&0\\
		0&H_D^*(-{\bf k})
	\end{array}
	\right),\nonumber\\
	&&H_D({\bf k})=\epsilon({\bf k})+d_i({\bf k})\sigma_i.
	\label{eq:H0}
\end{eqnarray}
where
\begin{eqnarray}
	&&d_1+id_2=\mathcal{A}(k_x - ik_y)=\mathcal{A}k_-,\nonumber\\
	&&d_3=\mathcal{M}-\mathcal{B}(k_x^2+k_y^2),\nonumber\\
	&&\epsilon=\mathcal{C}-\mathcal{D}(k_x^2+k_y^2).
	\label{eq:para1}
\end{eqnarray}
The two components of the Pauli matrices $\sigma$ denote the $E_1$ and $H_1$
sub-bands, while the two diagonal blocks $H_D({\bf k})$ and $H_D^*(-{\bf k})$
of $H_{eff}$ represent spin-up and spin-down states, related to
each other by time reversal symmetry. At the critical thickness,
the relativistic mass $\mathcal{M}$ in (\ref{eq:para1}) equals to zero.
If we then only keep the terms up to linear order in $\bf k$
for each spin, $H_D({\bf k})$ or $H_D^*(-{\bf k})$ correspond to
massless Dirac Hamiltonians. A HgTe quantum well at $d=d_c$ is thus a direct
solid state realization of a massless Dirac Hamiltonian. Since it does not have any valley degeneracy,
a $d=d_c$ HgTe/(Hg,Cd)Te quantum well is, in a sense, half-graphene.
Besides the linear term forming the Dirac Hamiltonian,
there are additional effects in the HgTe Dirac system, such as the quadratic terms
in (\ref{eq:H0}), and the presence of Zeeman- and inversion asymmetry-induced terms,
which are discussed in detail in the supplementary material.

As explained in the introduction, a two-cone Dirac system is the simplest possible realization of Dirac fermions for
any two dimensional quantum well or thin film, which makes HgTe a
very interesting model system to investigate Dirac fermion physics.
Other benefits include the very high mobility (up to $1.5 \times 10^6$
cm$^2$/Vs for high carrier densities) and the possibility to study
the effects of a finite relativistic mass $\mathcal{M}$ (with both positive
and negative sign). In this paper, we describe magneto transport
experiments on gated zero gap HgTe wells that clearly demonstrate
the Dirac fermion physics expected from Eqs. (\ref{eq:H0}) and (\ref{eq:para1}).

\section{Experimental}
For these studies, we have grown by molecular beam epitaxy a number of modulation-doped HgTe/$\text{Hg}_{0.3}$$\text{Cd}_{0.7}$Te
quantum well structures on lattice-matched (Cd,Zn)Te substrates, with a nominal well width ranging  from
5.0 to 7.5 nm (yielding various relativistic masses $\mathcal{M}$), including several samples aiming for the critical thickness of 6.3 nm. From Fig. \ref{fig:Dirac} (a),
the reader can infer that the series includes both normal and inverted band gap structures. From X-ray reflectivity measurements on our quantum well structures\cite{stahl2010} we infer the existence of thickness fluctuations of the order of a monolayer in the samples, which corresponds to fluctuations in $\mathcal{M}$ of around 1 meV.
Subsequently, the wafers have been processed into Hall bar devices with dimensions
(length $L$ $\times$ width $W$) of (600 $\times$ 200) and (20.0 $\times$ 13.3)
$\mu$ m$^{2}$ using a low temperature positive optical
lithography process.
For gating purposes a 100 nm thick $\text{Si}_3\text{N}_4/\text{SiO}_2$
multilayer gate insulator and a 5/50~nm Ti/Au gate electrode are deposited.
Ohmic contacts are made by thermal indium bonding.
A micrograph of such a Hall bar device is shown in the inset of Fig.\ref{fig:bcross} (a).
At zero gate voltage, the devices are n-type conducting with carrier concentrations
around $5 \times 10^{11}~\text{cm}^{-2}$
and mobilities of several $\text{10}^5~\text{cm}^2 V^{-1}s^{-1}$ .

Transport measurements are carried out in a variable temperature magneto-cryostat at a temperature of 4.2~K,
unless indicated otherwise. Typically, a bias voltage of up to 10 mV is applied between current
contacts 1 and 6 (as denoted in the inset of Fig. \ref{fig:bcross}(a) ), resulting in a current I of approximately
1 $\mu$ A , as determined by measuring the voltage drop across a reference resistor in series with the sample.
The resulting longitudinal  ($V_{xx}$, contacts 3 and 5) and transverse ($V_{xy}$, contacts 2 and 3) voltages are detected simultaneously
yielding the  longitudinal ($\rho_{xx}=V_{xx}/I \times W/L$) and transverse ($\rho_{xy}=V_{xy}/I$) resistivities.

Applying a gate voltage $\text{V}_{\rm{G}}$ between the top gate and the 2DEG, the electron density
(and thus the Fermi energy) can be adjusted. As reported previously,
the carrier type can be varied from n-type conductance for positive
$\text{V}_{\rm{G}}$ to p-type behavior for negative $\text{V}_{\rm{G}}$. Hysteresis effects due to interfacial states
\cite{hinz2006} restrict the usable range of gate voltages to $\left|\text{V}_{\rm{G}}\right| < 4$~V. For reasons of comparison, we have adjusted the gate voltage axes in Figs. 2, 3 and 4
 such that $\text{V}_{\rm{G}} -\text{V}_{\rm{Dirac}}=0$ V corresponds to the Dirac point. $\text{V}_{\rm{Dirac}}$ varies from cool-down to cool-down, but typically is of order -1.2 V.

\section{Quantum Hall effect and the identification of zero-gap samples}

\begin{figure}
	\centering
	\includegraphics[width=0.6\textwidth]{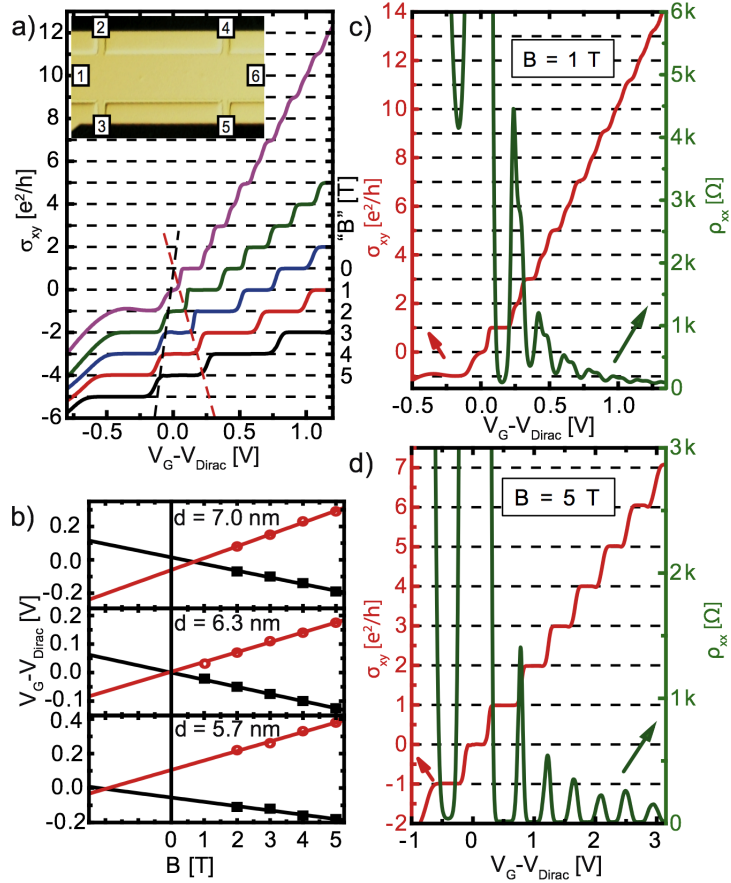}
	\caption{ (a) Gate voltage dependence of the transverse conductivity for a HgTe quantum well
	with a well thickness $d\approx$ 6.3 nm for fields of (from top to bottom) 1 (purple), 2 (green), 3 (blue), 4 (red), and 5 (black) Tesla. Note that the conductivity axis belongs to the 1 Tesla trace. All traces for higher magnetic fields have been shifted down by one conductance quantum per Tesla so as to show the determination of  $B^c_{\perp}$, using the "B"-axis on the right-hand ordinate as field axis (as based on Eq. \ref{eq:E0}, see text). Inset: Micrograph of a typical sample.
    (b) Determination of  $B^c_{\perp}$ for three exemplary quantum wells grown close to $\text{d}_{c}$. By extrapolating the crossing of the zero level spin states to the gate voltage needed to reach the Dirac point one finds $B^c_{\perp}$; when
     $B^c_{\perp} = 0 $ the well has the critical thickness $d_c \approx 6.3$ nm. See text for details.  (c,d) The transverse
    conductivity and longitudinal resistivity of a sample with well thickness $d\approx 6.3$ nm at 1 and 5 Tesla, respectively. }
	\label{fig:bcross}
\end{figure}

In Fig. \ref{fig:bcross}(a) we plot the Hall conductivity
$\sigma_{xy}=\rho_{xy}/(\rho^2_{xy}+\rho^2_{xx})$ at various fixed magnetic fields for a sample with $ d \approx d_c \simeq 6.3$ nm as a function of the gate voltage. The conductivity axis is correct for the trace taken at 1 T, while the traces for higher fields
have been offset by a constant amount (in this case one conductance quantum) per Tesla, for reasons that will become obvious shortly. First, we note that the traces show well developed quantum Hall plateaus, even for fields as low as 1 T. At this low field, the spin-derived Hall-plateaus (the conductance plateaus at an even integer times $\frac{e^2}{h}$) are still less broad than the orbital-induced ones (plateaus at an odd integer times $\frac{e^2}{h}$), which facilitates their assignment. Obviously, because of the large g-factor of HgTe ($g^*=55.5$ for this well, see below) the Landau levels are always spin-resolved. A full assessment of Dirac behavior will thus have to come from the field and energy dependence of the Landau level structure, which we will provide below.

First, we will address another question - is the sample really zero gap?
Since MBE growth calibration is not sufficiently precise to consistently grow a quantum well of exact critical thickness, we require another independent means to assess the well thickness.
We have found a simple procedure by analyzing the quantum Hall data of our samples. Specifically, it turns out that the crossing point of the lowest Landau levels for the electron and heavy-hole sub-bands is a precise measure of well thickness.
By solving the Landau levels of the effective Hamiltonian
(\ref{eq:H0}) in a magnetic field, we find that each of the spin blocks exhibits
a 'zero mode' (n=0 Landau level), which is one of the important differences between Landau levels
of materials described by a Dirac Hamiltonian and
those of more traditional metals\cite{jackiw1984}. The energy of the zero mode
is given by
\begin{eqnarray}
	&&E^\uparrow_0=\mathcal{C}+\mathcal{M}-\frac{eB_\perp}{\hbar}(\mathcal{D}+\mathcal{B}), \nonumber\\
	&&E^\downarrow_0=\mathcal{C}-\mathcal{M}+\frac{eB_\perp}{\hbar}(-\mathcal{D}+\mathcal{B})
    \label{eq:E0}
\end{eqnarray}
for the spin-up and spin-down block, respectively.
Here $B_{\perp}$ is the perpendicular magnetic field.
The spin splitting, given by $E^{\uparrow}_0-E^{\downarrow}_0=2\mathcal{M}-2\mathcal{B}\frac{eB_{\perp}}{\hbar}$,
thus increases linearly with magnetic field.
From (\ref{eq:E0}), we find that there is a critical
magnetic field $B^c_{\perp}=\frac{\hbar \mathcal{M}}{e\mathcal{B}}$, where
the two zero mode spin levels become degenerate, $E^\uparrow_0=E^\downarrow_0$.
In the inverted regime $\mathcal{M}/\mathcal{B}>0$, this degeneracy occurs at a positive magnetic field $B^c_\perp>0$, while in the normal regime where $\mathcal{M}/\mathcal{B}<0$, the crossing extrapolates to a negative value of $B^c_\perp<0$. For a well exactly at the critical thickness $d_c$ we have $\mathcal{M}/\mathcal{B}=0$, and the crossing point
will occur at zero field, $B^c_\perp=0$. Therefore the position of
the crossing point of the spin states of the lowest electron and hole Landau levels
at zero magnetic field will give us a direct indication
for the existence of a Dirac point ($\mathcal{M}=0$) in the quantum well.

Applying this procedure to the experimental data of Fig. \ref{fig:bcross} (a) (and similar data from the other quantum wells in our growth series) is straightforward. Since the Landau levels are already well defined at small magnetic fields, we can easily identify the $\pm 1$ Landau levels corresponding to the two spin blocks of the zero mode as
the boundaries of the $\sigma_{xy}=0$ plateau, at various magnetic fields. The constant offset between the different plots in Fig. \ref{fig:bcross} (a) implies that we can now translate the vertical axis into a field axis with a spacing of 1 T between the scans (the "B"-axis in the figure) and we can directly plot the linear spin splitting predicted by Eq. (\ref{eq:E0}) in Fig. \ref{fig:bcross} (b). Extrapolating the linear behavior in the graph allows us to determine $B^c_{\perp}$, which in this case leads to $B^c_{\perp} \approx 0 \;$ T - this sample has a Dirac mass close to zero.

As an illustration of the efficiency and sensitivity of this procedure, Fig. \ref{fig:bcross} (b) shows the extraction of $B^c_{\perp}$ for three different samples. The sample in the upper panel has an inverted band structure since $B^c_{\perp}>0$ (from a more detailed fit we find $ d = 7.0$ nm). The
middle panel corresponds to the data of Fig. \ref{fig:bcross} (a), where the intersection is at $B^c_{\perp} \approx 0~T$, corresponding to $\mathcal{M}=0$, and finally the sample in the bottom panel has a not-inverted band structure since the crossing point occurs for $B^c_{\perp}<0~T$ (and corresponds to a well-width of approximately 5.7 nm).

\section{Further characterization of a zero-gap sample}
In the following, the sample with $B^c_{\perp} = 0$ T of Fig.\ref{fig:bcross} (b)  is further investigated. Figs. \ref{fig:bcross} (c) and (d) show the Hall conductivity
of this sample at 1 and 5 T, respectively, in combination with the Shubnikov-de Haas oscillations in the longitudinal resistance.
The first thing to note is the quantization of the Hall plateaus. Orbital quantization yields
plateaus at odd multiples of $e^2/h$, with additional even-integer plateaus due to spin splitting already observable at 1 T. This is the unusual ordering of the Hall plateaus that results from the Dirac Hamiltonian\cite{castro2009}. Moreover, the observed plateaus occur at one half the conductance of the plateaus observed for graphene\cite{novoselov2005,zhang2005} - a direct consequence of the fact that the HgTe quantum well only has a single (spin degenerate) Dirac cone, where graphene has two. Furthermore, we always observe a plateau at zero conductance in the Hall traces, which is different from the low-field behavior in graphene\cite{castro2009}. The zero conductance Hall plateau is always accompanied by a quite large longitudinal resistivity, which is once more an indication that - already at 1T - the sample is gapped due to spin splitting.

\begin{figure}
	\centering
	\includegraphics[width=0.9\textwidth]{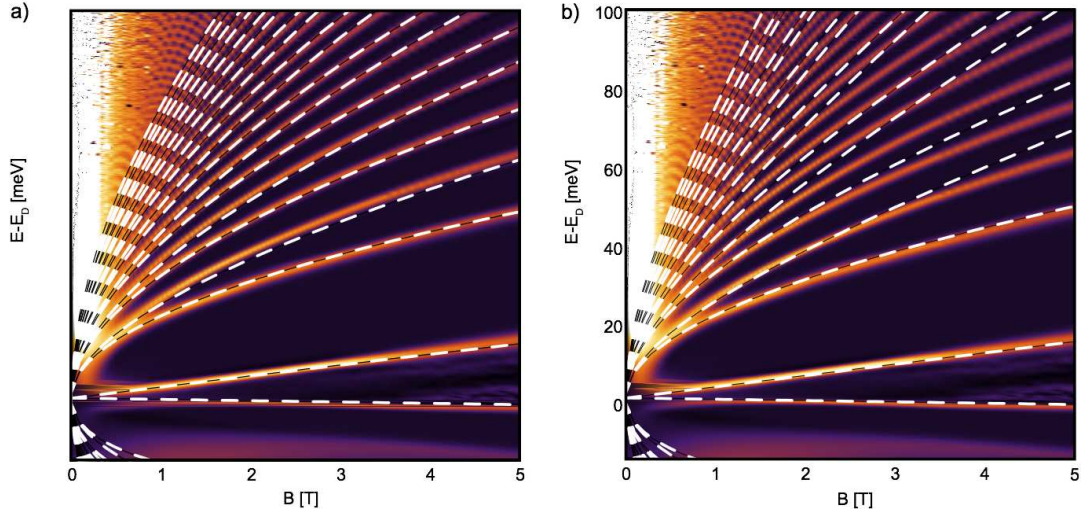} 
	\caption{Experimental Landau level fan charts obtained by plotting $ \partial\sigma_{xy} / \partial{\text{V}_{\rm{G}}} $ in a color-coded 3-dimensional graph as a function of both $\text{V}_{\rm{G}}$ and $B_{\perp}$. Energies are measured with respect to the Dirac point. In (a) this data is compared with the results of a calculation (dashed black-and-white line) from our 8-band $k \cdot p$ model. (b) Comparison with the fan chart computed from the Dirac Hamiltonian (Eq. \ref{eq:H0}) (dashed black-and-white line) using the parameters described in the text.}
	\label{fig:chart3er}
\end{figure}

To further validate our claim that this sample boasts a zero gap Dirac Hamiltonian at low energies, we  plot in Fig.\ref{fig:chart3er} (a) a Landau level fan chart.
This chart was obtained by plotting the derivative $ \partial\sigma_{xy} / \partial{\text{V}_{\rm{G}}} $ in a color-coded 3-dimensional graph as a function of both $\text{V}_{\rm{G}}$ and $B_{\perp}$. When the sample exhibits a quantum Hall plateau, the Hall conductance obviously is constant and its derivative is zero; when a Landau level crosses the Fermi energy,  $\partial\sigma_{xy} / \partial{\text{V}_{\rm{G}}}$ reaches a maximum,
which can be conveniently indicated by the color coding.
To translate the gate voltage axis to an energy scale for the band structure, we assume that the gate acts as a plane capacitor plate, and calculate the electron density in the quantum well as a function of energy using our 8-band $k \cdot p$ model\cite{novik2005}, assuming the well has the critical thickness $d= 6.3 \;$ nm. Furthermore, in the supplementary material, we calculate the density of states as a function of magnetic field for fixed electron density and compare the results with the experimental data on the Shubnikov-de Haas oscillations. The good agreement of the node position and spin splitting
between the experiment and theory verifies the validity of the 8-band $k \cdot p$ model.
The dashed white lines in Fig.\ref{fig:chart3er} (a) give the Landau level
dispersion predicted by our calculation; the very good agreement with the experimental peaks in $\partial\sigma_{xy} / \partial {\text{V}_{\rm{G}}}$  is evidence that our $\text{V}_{\rm{G}}$ to $E$ conversion is self-consistent.

The Landau-level dispersion in Fig. \ref{fig:chart3er} (a) shows
all the characteristics expected from our Dirac Hamiltonian (\ref{eq:H0}). Besides the zero mode of Eq.(\ref{eq:E0}), solving the Landau levels of the effective Hamiltonian
(\ref{eq:H0}) in a magnetic field, yields for
the higher Landau levels ($n=1,2,\cdots\cdots$) ($\mathcal{C,M}=0$):
\begin{eqnarray}
	&&E^\uparrow_\alpha(n)=-\frac{e}{\hbar}B_{\perp}(2 \mathcal{D} n + \mathcal{B}) \nonumber\\
	&&+\alpha \sqrt{2 n \mathcal{A}^2 \frac{e}{\hbar}B_{\perp}+ \left(\frac{e}{\hbar}B_{\perp}\right)^2(\mathcal{D}+2\mathcal{B}n)^2}\nonumber\\
	&&E^\downarrow_\alpha(n)=-\frac{e}{\hbar}B_{\perp}(2 \mathcal{D} n - \mathcal{B}) \nonumber\\
	&&+\alpha \sqrt{2 n \mathcal{A}^2 \frac{e}{\hbar}B_{\perp}+ \left(\frac{e}{\hbar}B_{\perp}\right)^2(\mathcal{D}-2\mathcal{B}n)^2}
    \label{eq:LL}
\end{eqnarray}
where $E^\uparrow_\alpha$, $E^\downarrow_\alpha$ refer to the two spin blocks of our Dirac Hamiltonian, Eq. (\ref{eq:H0}), and $\alpha=+,-$ denote the conduction and valence band, respectively.
With optimized parameters ( we use $\mathcal{C,M} = 0$  meV, $\mathcal{D} = -682$ meV$\cdot$nm$^2$, $\mathcal{B} = -857$ meV$\cdot$nm$^2$, $\mathcal{A}= 373$ meV$\cdot$nm )
the Landau level dispersion described by Eq. (\ref{eq:LL}) are plotted (dashed white lines) as a function of magnetic field in Fig. \ref{fig:chart3er} (b). Clearly, the Dirac model agrees well with our experiment for low magnetic field and low-index Landau levels, but gradually breaks down when the magnetic field is increased.

In the low magnetic field limit, one easily finds that Eq. (\ref{eq:LL}) (for the conduction band) reduces to $ E^{\uparrow(\downarrow)}_+(n) \approx \mathcal{A}\sqrt{2n\frac{e}{\hbar}B_{\perp}}
-(\frac{e}{\hbar}B_{\perp})(2\mathcal{D}n \pm \mathcal{B})  $ up to $B_\perp$ linear terms. This corresponds to
the square-root magnetic field dependence that has meanwhile become the signature of Dirac fermion behavior
in graphene\cite{castro2009}, with an additional linear term reflecting the large effective g-factor $g^*$ of the HgTe quantum well. Defining $\mu_Bg^*B_\perp=E^\uparrow_{+}(0)-E^\downarrow_+(0)$, we find $g^*\approx 55.5$.
There are two physical origins for the large $g^*$. Due to the zero gap nature of the present system, the most important
contribution comes from orbital effects which are fully incorporated in the Dirac Hamiltonian.
However, there is also a contribution from Zeeman-type terms,
which is not included in the Dirac Hamiltonian (\ref{eq:H0}). This term is less important
than the orbital part and will be discussed in the supplementary material.

Another effect that is not included in our model Dirac Hamiltonian is the inversion asymmetry of the system. In principle,
the HgTe quantum well has structural (SIA) and bulk inversion asymmetry (BIA) \cite{koenig2008,rothe2010}, both of which can couple Dirac cones with opposite spin.
From the node position of Shubnikov-de Haas oscillations (the data are presented in the supplementary material), we find that the spin splitting due to SIA is less than 2.5 meV at the largest experimentally accessible Fermi energy, decaying rapidly with density\cite{rothe2010}.
The present experiment does not show any evidence of the BIA term; a previous theoretical estimate shows
that the BIA term has an energy scale of about 1.6 meV\cite{koenig2008}. We conclude that also the SIA and BIA terms are small compared to the other terms in the Dirac Hamiltonian of Eq. (\ref{eq:H0}). Moreover, they cannot cause the opening of a gap in the quantum well spectrum. The relevance of SIA and BIA terms is discussed in more detail in the supplementary material.

\section{Zero field behavior}

\begin{figure}
	\centering
	\includegraphics[width=0.7\textwidth]{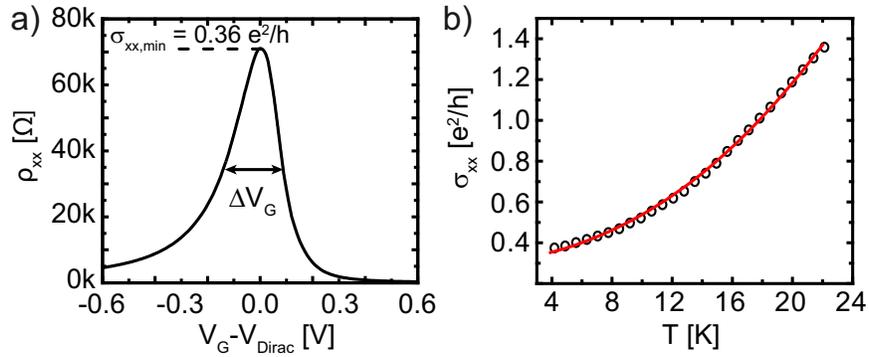}
	\caption{Resistivity $\rho_{xx}$ vs. gate voltage $\text{V}_{\rm{G}}$ (Dirac-peak) measured on a zero gap quantum well in the absence of an external magnetic field. This trace was taken at a sample temperature of 4.2 K. (b) Conductivity at  the Dirac point as a function of temperature. Open circles are experimental data, the red line is a fit to Eqs. (5) and (6). }
	\label{fig:sxx}
\end{figure}

Having thus established that we indeed can describe our quantum well as a zero gap Dirac system, we now turn to its characteristics at zero magnetic field. Fig. \ref{fig:sxx} (a) plots the resistivity $\rho_{xx}$ vs. gate voltage, often called the Dirac-peak in the graphene community, in this limit. The graph clearly shows the expected peaked
resistivity and exhibits an asymmetry between n- and p-regime which can be attributed to the large hole mass (increased density of states). In graphene, the width of the Dirac-peak is often regarded as a measure of the quality of the sample \cite{bolotin2008}. The width of the Dirac peak in Fig. \ref{fig:sxx} (a) corresponds to a carrier depletion of about $\Delta n \approx 3.0 \times10^{10}$cm$^{-2}$, which is comparable with the situation found in suspended graphene.

At the Dirac point, we find a minimum conductivity of $\sigma_{xx,min}=0.36~e^2/h$ at 4.2 K. Its temperature dependence
is shown in Fig. \ref{fig:sxx} (b), which conveys an initially quadratic temperature dependence, that for temperatures above about 12 K turns linear.
The existence of a finite minimal conductivity at vanishing carrier density
is a topological (Berry phase) manifestation of the conical singularity of the Dirac bands at ${\bf k}=0$.
Therefore, our observation of a minimal conductivity in HgTe quantum wells
provides independent evidence for the Dirac fermion behavior in this material.
The observed minimal conductivity (close to $e^2/\pi h$\cite{Fradkin86b,twor2006}) and the crossover
from quadratic ($\propto T^2$)  to linear ($\propto T$) increase with temperature can be understood from
calculations based on the Kubo formula, in which the current-current correlation function is evaluated for the effective Dirac Hamiltonian of Eq. (\ref{eq:H0}), assuming the presence of both well width fluctuations and potential disorder and including only the dominant terms linear in ${\bf k}$.
The details of these calculations are described in the supplementary material. Qualitatively, the temperature dependence of $\sigma_{xx}$ is

\begin{eqnarray}
&&
\sigma_{xx}\approx \frac{2}{\pi} \frac{e^2}{h}\frac{1}{1 +  \langle \mathcal{M}^2 \rangle/\Gamma^2 }
\nonumber
\\
&&
+ O\left( \frac{e^2}{h}\frac{k^2_{_B} T^2}{\Gamma^2} \right),
\, k_{_B}T\ll\Gamma,
\label{Sigma1}\\
&&
\sigma_{xx} \propto \frac{e^2}{h}\frac{k_{_B}T}{\Gamma},
\quad k_{_B}T\geq \Gamma,
\label{Sigma2}
\end{eqnarray}

where $\Gamma$ is the spectral broadening induced by spin-independent potential disorder.
In Eq.~(\ref{Sigma1}) the factor of 2 accounts for the spin degeneracy and
$\langle \mathcal{M}^2 \rangle\propto\langle (d-d_c)^2 \rangle$ is the variance of the gap
due to spatial deviations of the thickness $d$ from the critical value $d_c$.
From the X-ray reflectvity data on our samples\cite{stahl2010}, we estimate $\sqrt{ \langle \mathcal{M}^2 \rangle } \sim 1$ meV. This is comparable with typical values of $\Gamma$ estimated from the self-consistent Born approximation (see supplementary material). Therefore, for $\sqrt{ \langle \mathcal{M}^2 \rangle } \sim \Gamma$ the zero-temperature conductivity is approximately $\sigma_{xx}(T\to 0)\approx e^2/\pi h$, in agreement with the data.
The $T^2$ correction reflects the spectral smearing at energies below $\Gamma$.
In contrast, at $k_{_B}T\geq \Gamma$ the linear T-dependence of $\sigma_{xx}$ (\ref{Sigma2})
reflects the linear density of states, which is another manifestation of the Dirac fermion physics
in HgTe quantum wells.

In conclusion, our paper reports the first experimental discovery of a two dimensional massless
Dirac fermion in a single valley system. The high mobility in the HgTe quantum wells should allow us to directly study ballistic transport phenomena that so far have been hard to access for Dirac fermions\cite{du2008}. Moreover, the material offers an additional parameter for the experiments in that the effects of a finite Dirac mass can now be studied in detail.

{\it Acknowledgements.} We acknowledge useful discussions with C. Gould and X.L. Qi
	 and thank E. Rupp and F. Gerhard for assistance in sample growth
	 and in the transport experiments. This work was supported
	 by the German Research Foundation DFG
	 (SPP 1285 'Halbleiter Spintronik', DFG-JST joint research program,
	 Emmy Noether program (P.R.) and grants AS327/2-1 (E.G.N.),
	 the Alexander von Humboldt Foundation (C.X.L. and S.C.Z.) and the US Department of Energy,
	 Office of Basic Energy Sciences, Division of Materials Sciences and Engineering,
	 under contract DE-AC02-76SF00515 (S.C.Z).

\section{Supplementary online material}
In the supplementary material , we will give theoretical details connected  with the analysis of the effective $g$ factor, Shubnikov de Haas oscillations and minimum conductivity in HgTe quantum wells.

\subsection{Effective Hamiltonian for HgTe quantum wells}
In this section, we will discuss the complete expression of the effective Hamiltonian
of HgTe quantum wells near the critical thickness $d_c$. As first described by Bernevig,
Hughes and Zhang\cite{bernevig2006d}, the low energy physics of HgTe quantum wells
is determined by four states $|E_1,\frac{1}{2}\rangle$,
$|H_1,\frac{3}{2}\rangle$, $|E_1,-\frac{1}{2}\rangle$ and $|H_1,-\frac{3}{2}\rangle$.
With these four states as basis, the complete Hamiltonian of the system when a magnetic field $B_\perp$ is applied in the the z-direction can be written as
\begin{eqnarray}
    \hat{H}=H_{eff}+H_{Zeeman}+H_{SIA}+H_{BIA}
    \label{eq:Sup_Ham}
\end{eqnarray}
The effective Hamiltonian is given by Eq. (1) in the main part of the article, where a Peierls substitution
$\bold{k}\rightarrow \bold{k}+\frac{e}{\hbar}\bold{A}$ has been applied ($\bold{A}$ is
the magnetic vector potential). All parameters $\mathcal{A}$,
$\mathcal{B}$, $\mathcal{C}$, $\mathcal{D}$ and $\mathcal{M}$ can be determined
by fitting to the experimental Landau level dispersion at $T=4.2$ K for
$n_{2DEG}\rightarrow0$ (we neglect in our calculations
the dependence of the parameters on the electron density), which are listed in the table
~\ref{Parameters}.

The Zeeman term $H_{Zeeman}$ has the form\cite{koenig2008}

\begin{eqnarray}
    H_{Zee}=\frac{\mu_BB_\perp}{2}\left(
    \begin{array}{cccc}
        g_E&&&\\
        &g_H&&\\
        &&-g_E&\\
        &&&-g_H
    \end{array}
    \right)
    \label{eq:Sup_HZee}
\end{eqnarray}

with $\mu_B=\frac{e\hbar}{2m_0}$ and the four band effective g-factor $g_E$ ($g_H$) for
$E_1$ ($H_1$) bands.

Since in HgTe quantum wells the inversion symmetry is broken, we also need to discuss
two terms that result from inversion asymmetry, i.e. the structural inversion asymmetry (SIA) term and the bulk inversion asymmetry (BIA) term. The SIA term is due
to the asymmetry of the quantum well potential and has the form\cite{rothe2010}

\begin{eqnarray}
    H_{SIA}=\left(\begin{array}{cccc}0&0&i\xi_ek_-&-i\chi k^2_-\\0&0&i\chi k^2_-&i\xi_hk^3_-\\
    -i\xi_e k_+&i\chi k^2_+&0&0\\-i\chi k^2_+&-i\xi_hk^3_+&0&0\end{array}\right).
    \label{eq:Sup_HSIA}
\end{eqnarray}

In (001) grown HgTe quantum wells,, the Rashba spin splitting is dominantly responsible
for the beating pattern of the Shubnikov-de Haas oscillations. By comparing
the results of a Kane model calculation with the experimental data, one can thus directly determine the
Rashba coefficient. As discussed in the following
section, we find that for a quite large range of gate voltages, the Rashba
spin splitting is less than 2.5 meV near the Fermi energy, which corresponds to
$\xi_e\approx 16$ meV$\cdot$nm, $\chi\approx 2.0$ meV$\cdot$nm$^2$ and $\xi_h\approx
5.0$ meV$\cdot$nm$^3$. Furthermore we find that the electron Rashba splitting $\xi_e$ is always
dominant over the other two terms.

Since the zinc-blend crystal structure of HgTe is not inversion-symmetric, additional BIA terms
appear in the effective Hamiltonian, given by\cite{koenig2008}
\begin{eqnarray}
    H_{BIA}=\left(\begin{array}{cccc}0&0&0&-\Delta_0\\0&0&\Delta_0&0\\
    0&\Delta_0&0&0\\-\Delta_0&0&0&0\end{array}\right).
    \label{eq:Sup_HBIA}
\end{eqnarray}
Since the BIA Hamiltonian is a constant term, it will only change the the Dirac point to a circle
and the system remains gapless. An early estimate of magnitude of BIA term gives $\Delta_0\approx1.6$meV\cite{koenig2008},
which is of the same order of magnitude as the disorder broadening and the fluctuations of the band gap due to variations in well width. Therefore in the present experiment we do not find evidence of a pronounced effect from this term.

\begin{table}[htb]
  \centering
  \begin{minipage}[t]{0.8\linewidth}
    \caption{\label{Parameters} The parameters used in the effective model.
      We take $\mathcal{C}$ to be zero to shift the Dirac point to zero energy. }
\label{tab:parameter}
\begin{tabular}{|c|c|}
    \hline
      $\mathcal{A}({\rm meV\cdot nm})$    & 373 \\\hline
      $\mathcal{B}({\rm meV\cdot nm})$  & -857 \\\hline
      $\mathcal{D}({\rm meV})$             & -682 \\\hline
      $\mathcal{M}({\rm meV})$             & -0.035 \\\hline
      $\Delta_0({\rm meV})$                & 1.6 \\\hline
      $g_{E}$                             & 18.5 \\\hline
      $g_{H}$                             & 2.4 \\
    \hline
  \end{tabular}%
  \end{minipage}
\end{table}

Neglecting the SIA and BIA terms, the Landau level spectrum is described by
\begin{eqnarray}
    &&E^\uparrow_\alpha(n)=  -\frac{eB_{\perp}}{\hbar}(2\mathcal{D}n + \mathcal{B})+\frac{\mu_B B_{\perp}}{4}( g_E+g_H)\nonumber\\
    &&+\alpha \sqrt{2 n \mathcal{A}^2 \frac{eB_{\perp}}{\hbar}+ \left(\mathcal{M}-B_{\perp}\left(\frac{e}{\hbar}(\mathcal{D}+2\mathcal{B}n)-\frac{\mu_B }{4}( g_E-g_H)\right)\right)^2}\nonumber\\
    &&E^\downarrow_\alpha(n)=-\frac{eB_{\perp}}{\hbar}(2\mathcal{D}n - \mathcal{B})-\frac{\mu_B B_{\perp}}{4}( g_E+g_H)\nonumber\\
    &&+\alpha \sqrt{2 n \mathcal{A}^2 \frac{eB_{\perp}}{\hbar}+ \left(\mathcal{M}-B_{\perp}\left(\frac{e}{\hbar}(-\mathcal{D}+2\mathcal{B}n)+\frac{\mu_B }{4}( g_E-g_H)\right)\right)^2},\nonumber\\
    \label{LLspectrum}
\end{eqnarray}
where $n=1,2,...$, and $\alpha=+$ ($\alpha=-$) for the conduction (valence band)
and the parameter $\mathcal{C}$ is taken to be zero, setting the Dirac point at zero energy.
The zero mode states ($n=0$) have the dispersion:

\begin{eqnarray}
    &&E^\uparrow_0=   \mathcal{M}-\frac{eB_{\perp}}{\hbar}( \mathcal{D}+\mathcal{B})+\frac{\mu_B B_{\perp}}{2}g_E
    \nonumber\\
    &&E^\downarrow_0=-\mathcal{M}+\frac{eB_{\perp}}{\hbar}(-\mathcal{D}+\mathcal{B})-\frac{\mu_B B_{\perp}}{2}g_H,
    \label{LLspectrum0}
\end{eqnarray}
and the zero mode splitting is:

\begin{equation}
    \Delta E_s=2 \mathcal{M}-\frac{2eB_{\perp}}{\hbar}\mathcal{B}+\frac{\mu_B B_{\perp}}{2}(g_E+g_H).
    \label{Splitting}
\end{equation}

Finally, we note that the total spin splitting, defined
as the energy difference between the spin-up and spin-down zero modes, has
two origins in the four band effective model ($\mathcal{M}\rightarrow0$).
One comes from the Zeeman term, which gives the energy splitting
$\Delta E_{s2}=\mu_Bg^*_{2}B_\perp=\frac{\mu_BB_\perp}{2}\left( g_E+g_H \right)$
with $g^*_2=10.5$. The other origin stems from
the combined orbital effects of the linear and quadratic terms in the Hamitonian $H_{eff}$,
and is given by $\Delta E_{s1}=\mu_{B} B_{\perp} g^*_1=-\frac{2eB_\perp}{\hbar}\mathcal{B}$
with $g^*_1=45$. These two terms
together give the effective g-factor $g^*$ defined in the main part of the article.

\subsection{Shubnikov-de Haas oscillations}
\begin{figure} [h]
\begin{center}
\includegraphics[width=0.95\columnwidth]{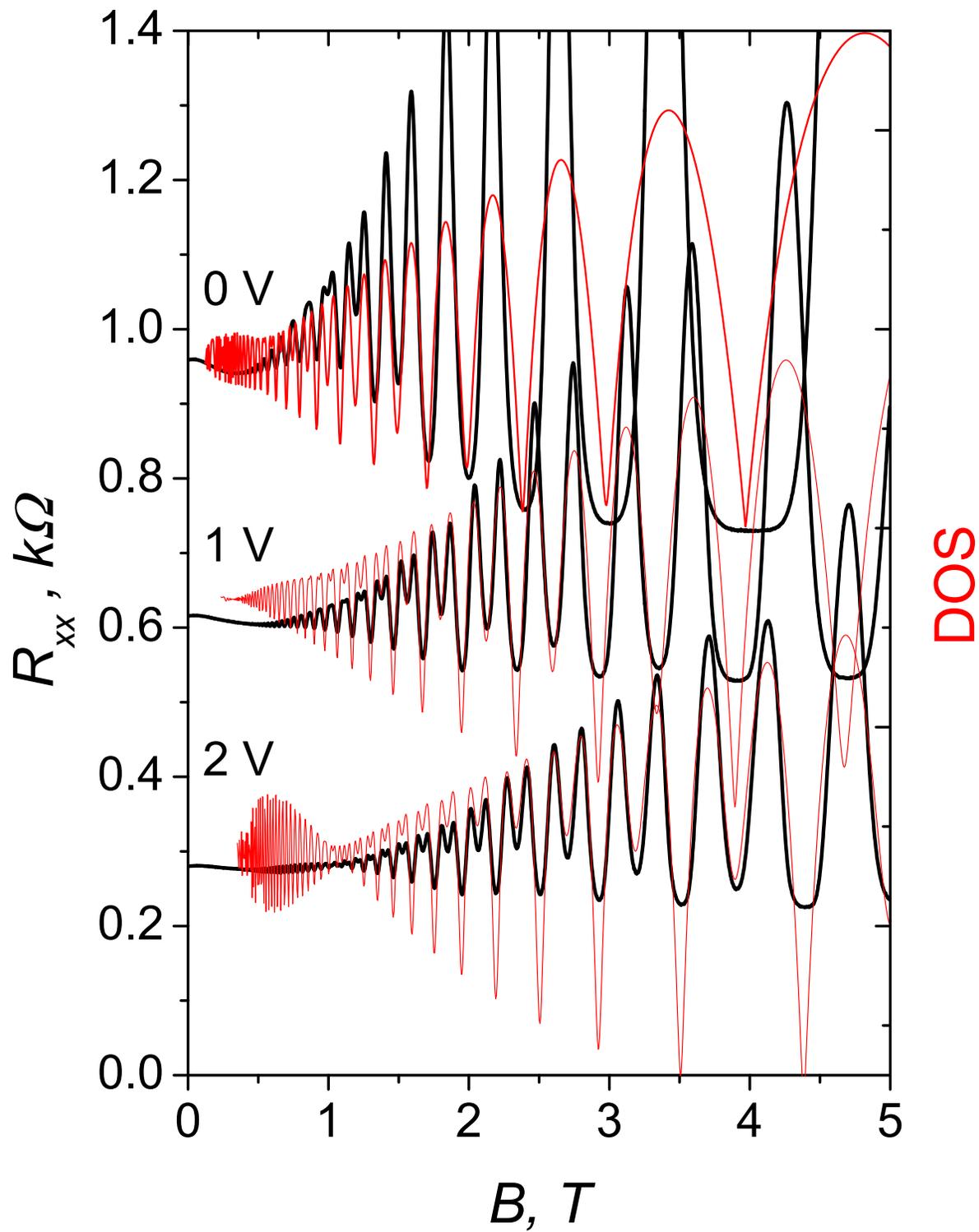}
\end{center}
\caption{\label{DOS} Calculated Fermi level density of states of a $d=d_c$ HgTe quantum well for various positive gate voltages as a function of perpendicular magnetic field (thin red lines) compared with the
experimental SdH oscillations (black thick lines).}
\end{figure}

In order to compare the Kane-model calculations with the experimental data,
the density of states (DOS) at the Fermi level was
calculated from the Landau level spectrum (see Ref. \cite{novik2005}
for more details). The Shubnikov-de Haas (SdH) oscillations
observed in the experiments are directly related to the oscillations of the
DOS at the Fermi energy. In Fig.~\ref{DOS} the
calculated DOS, broadened by convolution with a Gaussian with a width
$\Gamma_{0}$=1.2~meV, is displayed together with SdH data for
three different values of the gate voltage. A very good agreement between experiment and theory
is evident. For all three different gate voltages,  we find that
there are always two sets of minima in the oscillations, one deep and one shallow,
which result from the two sets of Landau levels for opposite spin (cf. Eqs. (\ref{LLspectrum})
and (\ref{LLspectrum0})). We first make the obvious identification that the deep minima result from the Landau level splitting and the shallow minima from the spin splitting. It is now instructive to plot the inverse of the magnetic field value for the positions of the deep minima as a function of the number N
associated with the deep minima of the oscillation, as done previously to demonstrate the implications of the Berry phase of the Dirac Hamiltonian for the quantum Hall effect in graphene\cite{novoselov2005}.
As shown in Fig. 6, we find that when $\text{V}_{\rm{G}}=2$ V, a straight line fit to the data extrapolates to $N=0$, while for $\text{V}_{\rm{G}}=0$ V, the fit extrapolates to $N=1/2$. This different behavior can be understood from the effective Hamiltonian (1) in the main text of the article. For $\text{V}_{\rm{G}}=0$ V, the electron density is low and the Fermi energy is near the Dirac point. In this limit, the band dispersion is dominated by the linear term in wave vector $k$ and the deep minima correspond to the filling factors $\nu=2(N-\frac{1}{2})$ (the factor of 2 takes
the spin into account). The intercept of the straight-line fit evidently corresponds to the filling factor $\nu=0$,
which explains the  $N=1/2$ intercept. However, for $\text{V}_{\rm{G}}=2$ V, the electron
density is increased and the Fermi energy is far away from the Dirac point.
Consequently, the linear term is no longer dominant and other terms, such as the quadratic ones, will come into play. In this limit, the system recovers the usual behavior of a two dimensional electron gas and the deep minima correspond to the filling factors $\nu=2N$, implying that the intercept occurs at $N=0$. Note that in HgTe, in contrast with graphene,  one always has additional shallow minima besides the deep ones due to the coexistence of the linear term and other type of terms, such as quadratic or Zeeman terms.
Another important feature of the SdH oscillations in Fig.~\ref{DOS} is the appearance
of a beating pattern when $\text{V}_{\rm{G}}=2$V, indicating the occurrence of Rashba
spin splitting\cite{winklerbook2003}, which from this data is estimated to be 2.5 meV. The beating feature is not observable when the gate voltage is in the range of 0 $\sim$ 1 V, which indicates
that the system becomes more symmetric for low electron densities. An extensive discussion of this effect can be found in Ref. \cite{rothe2010}.

%
%

\subsection{Calculation of the minimal conductivity}

In this section of the supplementary material, we discuss details of the calculation of the minimal
conductivity given by Eqs. (5) and (6) in the main part of the article.

We use the Kubo formula for the longitudinal ($xx$) dc conductivity,

\begin{eqnarray}
\sigma_{xx}=2\times\pi e^2\hbar\int\,
d\epsilon\left(- \frac{df}{d\epsilon} \right)\int\frac{d^2{\bf k}}{(2\pi)^2}
{\rm Tr}
\left[
\hat{v}_x\, {\hat A}_{{\bf k},\epsilon} \, \hat{v}_x \, \hat{ A}_{{\bf k},\epsilon}
\right],
\label{Kubo}
\end{eqnarray}
\begin{eqnarray}
\hat{v}_x=\frac{1}{\hbar}\frac{ \partial H_D({\bf k}) }{\partial k_x}, \quad
\hat{ A }_{{\bf k},\epsilon}=\frac{ \hat{G}^{^A}_{{\bf k},\epsilon} - \hat{G}^{^R}_{{\bf k},\epsilon} }{2\pi i}.
\label{A}
\end{eqnarray}
Here the velocity operator $\hat{v}_x$, spectral function $\hat{ A }_{{\bf k},\epsilon}$ and
retarded/advanced Green's functions $\hat{G}^{^{R/A} }_{{\bf k},\epsilon}$ are
$2\times 2$ matrices in $E_1$-$H_1$ subband space of the effective Hamiltonian $H_D({\bf k})$
described by Eqs.~(1) and (2) of the main manuscript. We use the symbol (${\rm Tr}$) to designate the trace operation .
The spin degree of freedom is accounted for by the factor of 2 in Eq.~(\ref{Kubo}).
$f(\epsilon)$ is the Fermi function, $\epsilon$ is the energy measured from the neutrality point and
${\bf k}=(k_x, k_y, 0)$ is the wave vector in the plane of the quantum well (QW).

Our next step is to calculate the disorder-averaged Green's functions $\hat{G}^{^{R/A} }_{{\bf k},\epsilon}$.
In the minimal conductivity regime the most relevant types of disorder in our HgTe QWs are
the inhomogeneity of the carrier density and the spatial fluctuations of the QW thickness $d({\bf r})$ around
critical value $d_c$. The carrier density inhomogeneity induces random fluctuations of the electrostatic
potential in the QW, which we treat as weak gaussian disorder with standard averaging procedures
leading to the complex self-energy in the equation for $\hat{G}^{^{R/A} }_{{\bf k},\epsilon}$
(see, Eqs.~(\ref{Eq_G}) and (\ref{Eq_S}) below).
In this respect, we follow the previous theoretical work on graphene (see, e.g. Refs.~\cite{Shon98,Ostrovsky06}).
The new feature of our model is that it also accounts for the QW thickness fluctuations which is
a specific type of disorder in HgTe/CdTe structures.
This type of disorder induces sizable regions in the sample
where the effective Dirac mass ${\cal M}$ has small positive or negative values $\propto d({\bf r}) - d_c$:
\begin{equation}
 {\cal M}({\bf r})\approx {\cal M}^\prime \cdot (d({\bf r}) - d_c),
\label{M}
\end{equation}
where ${\cal M}^\prime\approx 15$ meV$\cdot$nm$^{-1}$ is a proportionality coefficient that we determine from band structure calculations.
We assume that the gap ${\cal M}({\bf r})$ varies slowly between the regions with ${\cal M}>0$ and ${\cal M}<0$
in the sense that the carrier motion {\em adiabatically} adjusts to the variation of ${\cal M}({\bf r})$.
Both ${\cal M}({\bf r})$ and its gradient $\nabla_{\bf r} {\cal M}({\bf r})$ are supposed to vanish upon
averaging over the whole sample area ($a$) so that the leading nonzero moment of ${\cal M}({\bf r})$ is the variance:

\begin{eqnarray}
\langle {\cal M}({\bf r})\rangle = \langle \nabla_{\bf r} {\cal M}({\bf r})\rangle =0,
\quad
\langle {\cal M}^2({\bf r})\rangle \not=0,
\quad
\langle ...\rangle
\equiv
a^{-1}\int d{\bf r}\, ...
\label{M_av}
\end{eqnarray}
In view of the adiabatic dependence ${\cal M}({\bf r})$ it is convenient to use the mixed representation for
the Green's functions defined by the Wigner transformation:

\begin{equation}
\hat{G}^{^{R/A} }_{ {\bf k} , \epsilon }({\bf r})
=\int \hat{G}^{^{R/A} }_\epsilon\left( {\bf r} + \frac{ {\bf r}_- }{2}, {\bf r} - \frac{ {\bf r}_- }{2} \right)
\,{\rm e}^{ -i\,{\bf k}\,{\bf r}_- }\, d{\bf r}_-,
\label{Wigner}
\end{equation}
where $\hat{G}^{^{R/A} }_{ {\bf k} , \epsilon }({\bf r})$ satisfies the equation:

\begin{eqnarray}
\left[
\epsilon + \mu - \Sigma^{^{R/A}}_\epsilon ({\bf r})  -
{\cal A}\mbox{\boldmath$\sigma$}\left(  {\bf k}  - \frac{i}{2} \nabla_{\bf r}  \right)
-
\sigma_z {\cal M}\left( {\bf r}  + \frac{i}{2} \nabla_{\bf k}  \right)
\right]
\hat{G}^{^{R/A}}_{ {\bf k} , \epsilon }({\bf r})
=
\hat{I}.
\label{Eq_G}
\end{eqnarray}
Here we omit the ${\bf k}^2$ corrections to the linear Dirac Hamiltonian [see Eqs.~(1) and (2) for $H_D({\bf k})$ in the main manuscript]
because the main contribution to the minimal conductivity comes from the vicinity of the ${\bf k}=0$ point
($\mu$ is the Fermi energy measured from the neutrality point).
For the same reason, in Eq.~(\ref{Eq_G}) the self-energy
$\Sigma^{^{R/A}}_\epsilon ({\bf r})$ (generated by the random potential fluctuations)
is taken at ${\bf k}=0$. It has been established earlier that the universal minimal conductivity follows already
from the self-consistent Born approximation or equivalent approaches
(e.g. Refs.~\cite{Fradkin86b,Ludwig94,Shon98,Ostrovsky06}).
We also adopt this approximation for the self-energy:

\begin{eqnarray}
\Sigma^{^{R/A} }_\epsilon ({\bf r}) =
\int \frac{ d^2{\bf q} }{(2\pi)^2}\, \zeta_{{\bf q}}\,
\hat{G}^{^{R/A} }_{ {\bf q} , \epsilon }({\bf r}),
\quad
\zeta_{-{\bf q}}=\zeta_{{\bf q}},
\label{Eq_S}
\end{eqnarray}
where $\zeta_{{\bf q}}$ is the Fourier transform of the correlation function of the random potential,
which is an even function of the wave-vector ${\bf q}$ due to the statistical homogeneity of the disorder.

In order to solve Eq.~(\ref{Eq_G}) we follow the same strategy as in the case of the uniform ${\cal M}$,
i.e. we first apply operator
$\epsilon + \mu - \Sigma^{^{R/A} }_\epsilon ({\bf r})  +
{\cal A}\mbox{\boldmath$\sigma$}\left(  {\bf k}  - \frac{i}{2} \nabla_{\bf r}  \right)
+ \sigma_z {\cal M}( {\bf r}  + \frac{i}{2} \nabla_{\bf k} )$
to both sides of the equation from the left.
Since ${\bf k}  - \frac{i}{2} \nabla_{\bf r}$ does not commute with
${\cal M}( {\bf r}  + \frac{i}{2} \nabla_{\bf k} )$ and $\Sigma^{^{R/A} }_\epsilon ({\bf r})$, there appear
additional gradient terms which are absent if ${\cal M}({\bf r})=const$:

\begin{eqnarray}
&
\left[
(\epsilon + \mu - \Sigma^{^{R/A} }_\epsilon ({\bf r}) )^2
- {\cal M}^2\left( {\bf r}  + \frac{i}{2} \nabla_{\bf k} \right)
- {\cal A}^2\left( {\bf k} - \frac{i}{2} \nabla_{\bf r}  \right)^2
\right.
&
\nonumber\\
&
\left.
+\frac{ {\cal A} }{2}\mbox{\boldmath$\sigma$}
\left(
i\nabla_{\bf r}\Sigma^{^{R/A} }_\epsilon ({\bf r}) +
{\bf z}\times\nabla_{\bf r}{\cal M}( {\bf r} ) +
{\bf z}\times\nabla_{\bf r}{\cal M}\left( {\bf r} + \frac{i}{2} \nabla_{\bf k} \right)
\right)
\right]
\hat{G}^{^{R/A} }_{ {\bf k} , \epsilon }({\bf r})
&
\label{Eq_G1}\\
&
=\epsilon + \mu - \Sigma^{^{R/A} }_\epsilon ({\bf r})   +
{\cal A} \mbox{\boldmath$\sigma$} {\bf k}   + \sigma_z {\cal M}( {\bf r}).
&
\nonumber
\end{eqnarray}
We are interested in the average Green's function over the sample area:
$\hat{G}^{^{R/A} }_{ {\bf k} , \epsilon }
\equiv a^{-1}\int \hat{G}^{^{R/A} }_{ {\bf k} , \epsilon }({\bf r})\, d{\bf r}$
and, respectively,
$\Sigma^{^{R/A} }_\epsilon
\equiv a^{-1}\int \Sigma^{^{R/A} }_\epsilon ({\bf r})\, d{\bf r}$.
Upon such averaging the linear terms ${\cal M}( {\bf r})$, $\nabla_{\bf r} {\cal M}( {\bf r} )$ and
$\nabla_{\bf r}\Sigma^{^{R/A} }_\epsilon ({\bf r})$ in Eq.~(\ref{Eq_G1}) vanish,
while the quadratic term ${\cal M}^2\left( {\bf r}  + \frac{i}{2} \nabla_{\bf k} \right)$ does not.
The latter is assumed to vary slowly in space, allowing us to
neglect the corrections $\propto \nabla_{\bf k}$ and
to obtain the following equations for the Green's function and the self-energy:

\begin{equation}
\hat{G}^{^{R/A} }_{ {\bf k} , \epsilon }=
\frac{ \epsilon + \mu - \Sigma^{^{R/A} }_\epsilon  + {\cal A} \mbox{\boldmath$\sigma$} {\bf k} }
{  (\epsilon + \mu - \Sigma^{^{R/A} }_\epsilon )^2  - \langle {\cal M}^2 \rangle - {\cal A}^2 {\bf k}^2 },
\Sigma^{^{R/A} }_\epsilon =
\int \frac{ d^2{\bf q}\, \zeta_{\bf q}}{ (2\pi)^2 }
\frac{ \epsilon + \mu - \Sigma^{^{R/A} }_\epsilon }
{  (\epsilon + \mu  - \Sigma^{^{R/A} }_\epsilon )^2  - \langle {\cal M}^2 \rangle  - {\cal A}^2 {\bf q}^2 }.
\label{G_S_av}
\end{equation}
These results are justified if the average slope of the gap variation,
$\sqrt{ \langle |\nabla_{\bf r} {\cal M} |^2\rangle}/\sqrt{ \langle {\cal M}^2 \rangle  }$, is small compared
to the characteristic electron wave-number $\Gamma_\epsilon/{\cal A}$,

\begin{equation}
 \frac{ \sqrt{ \langle |\nabla_{\bf r} {\cal M} |^2\rangle} }{\sqrt{  \langle {\cal M}^2 \rangle  }} \ll
 \frac{ \Gamma_\epsilon }{ {\cal A} },
 \quad
 \Gamma_\epsilon =|{\rm Im}\, \Sigma^{^{R/A} }_\epsilon|,
 \label{slow}
\end{equation}
where $\Gamma_\epsilon$ is the spectral broadening due to the finite elastic life-time.
It eliminates the infrared divergence of the ${\bf k}$ integral in Kubo formula (\ref{Kubo}),
providing an effective cut-off at small values of ${\bf k}$.

We now use Eq.~(\ref{G_S_av}) for $\hat{G}^{^{R/A} }_{ {\bf k},\epsilon }$
and $\hat{v}_x=\partial H_D({\bf k})/\hbar\partial k_x \approx ({\cal A}/\hbar)\sigma_x$
to calculate the ${\bf k}$ integral in Eq.~(\ref{Kubo}) and express the conductivity in the form:

\begin{equation}
\sigma_{xx}(T) = \int_{-\infty}^\infty \frac{d\epsilon}{ 4k_BT \cosh^2( \epsilon/2k_BT )  }\, \sigma_{xx}(\epsilon),
\label{sigma_T}
\end{equation}
\begin{eqnarray}
\sigma_{xx}(\epsilon)=\frac{2 e^2}{ \pi h }
&\times&
\frac{1}{2}
\left[
\frac{ \overline{\epsilon}^2 + \Gamma^2_\epsilon }{  2|\overline{\epsilon}|\Gamma_\epsilon  }
\arcsin\frac{ 2|\overline{\epsilon}|\Gamma_\epsilon}
{\sqrt{ (\Gamma^2_\epsilon - \overline{\epsilon}^2 + \langle {\cal M}^2 \rangle)^2 + 4\overline{\epsilon}^2\Gamma^2_\epsilon} }
\right.
\nonumber\\
& + &
\left.
\frac{ (\Gamma^2_\epsilon + \overline{\epsilon}^2)^2  + (\Gamma^2_\epsilon - \overline{\epsilon}^2)\langle {\cal M}^2 \rangle}
{ (\Gamma^2_\epsilon - \overline{\epsilon}^2 + \langle {\cal M}^2 \rangle)^2 + 4\overline{\epsilon}^2\Gamma^2_\epsilon }
\right],
\quad
\overline{\epsilon}=\epsilon + \mu - {\rm Re}\,\Sigma^{^{R/A} }_\epsilon.
\label{sigma_E}
\end{eqnarray}
For $\epsilon, \mu \to 0$ Eq.~(\ref{sigma_E}) yields the zero-temperature minimal conductivity $\sigma_{xx}(0)$.
To completely specify Eq.~(\ref{sigma_E}) we find the self-energy $\Sigma^{^{R/A} }_\epsilon$ from Eq.~(\ref{G_S_av}) in the form of a power-law expansion:
\begin{equation}
\Sigma^{^{R/A} }_\epsilon \approx C_0 + C_1\epsilon + C_2 \epsilon^2 + ...
\label{Expansion}
\end{equation}
Assuming $\zeta_{\bf q}=\zeta_0=const$ and a cutoff $\Delta={\cal A}^2/{\cal B}\approx 120$ meV
at high energies [where the quadratic term ${\cal B}k^2$ becomes comparable with the linear one
${\cal A}k$ in $H_D({\bf k})$], we obtain the expansion coefficients as\cite{footnotelimits}

\begin{eqnarray}
&
C_0=\mp i\Gamma,\quad \Gamma=\sqrt{  \Delta^2 e^{ -2/\alpha } - \langle {\cal M}^2 \rangle },
\quad \alpha=\zeta_0/2\pi{\cal A}^2,
&
\label{C0}\\
&
C_1=-\left(
\frac{\alpha\, \Gamma^2}{ \Gamma^2 + \langle {\cal M}^2 \rangle }
\mp \frac{i\mu}{\Gamma}
\right)^{-1},
C_2=\pm i
\frac{\alpha\, \Gamma (\Gamma^2 + 3\langle {\cal M}^2 \rangle) }{2(\Gamma^2 + \langle {\cal M}^2 \rangle)^2}
\, C^3_1,
\quad
\mu/\Gamma \ll \alpha \ll 1.
&
\label{C1C2}
\end{eqnarray}
This model adequately describes the observed minimal conductivity
and its temperature dependence. In particular,  at low temperatures $k_B T\ll \Gamma$
the $T$-dependence is quadratic:
\begin{equation}
\sigma_{xx}(T)\approx \frac{2 e^2}{ \pi h }
\left[
\frac{1}{1 + \langle M^2\rangle/\Gamma^2} +
\frac{1/9 + \langle M^2\rangle/\Gamma^2}{ (  1 + \langle M^2\rangle/\Gamma^2)^3 }\times
\frac{\pi^2 k^2_B T^2}{\Gamma^2}
\right].
\quad
\label{Sigma_min}
\end{equation}
At $k_BT\geq \Gamma$ the conductivity becomes approximately linear as a manifestation
of the linear density of states of the 2D Dirac fermions in HgTe QWs.
The fit to the experimental curve $\sigma_{xx}(T)$ in Fig.~4b
is achieved for $\sqrt{\langle M^2\rangle}=\Gamma=1$ meV and $\mu/\alpha^3\Gamma=0.06$.


\newpage

\begin{figure} [h]
\begin{center}
\includegraphics[width=0.95\columnwidth]{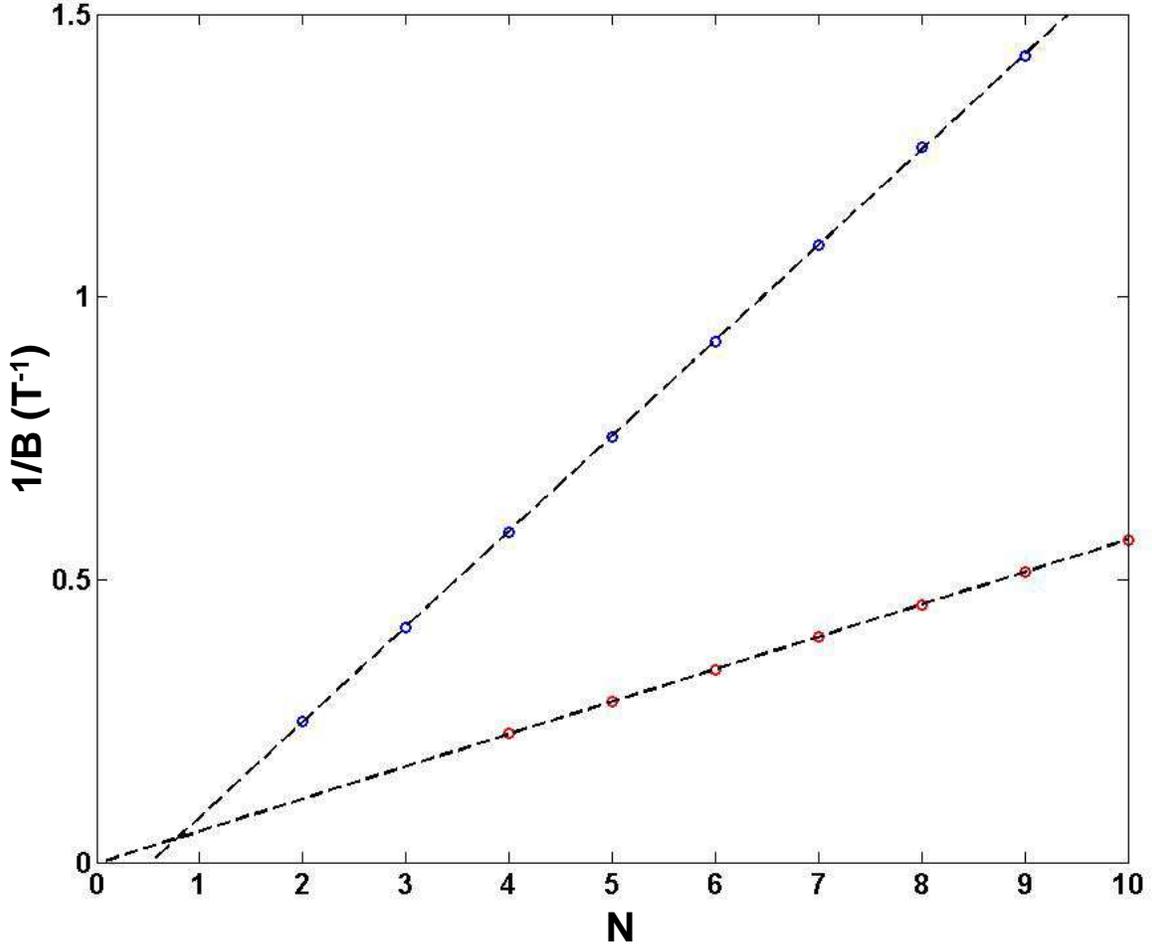}
\end{center}
\caption{\label{minimal} The inverse of the magnetic field at which
the deep minima in the Shubnikov-de Haas oscillations occur, as a function of ordinal number N
for $\text{V}_{\rm{G}}=2$ V (red circle) and $\text{V}_{\rm{G}}=0$ V (blue circle). The ordinal number N can also be related to the conductivity at the corresponding Hall plateau. For $\text{V}_{\rm{G}}=0$ V, the corresponding Hall conductivity for deep minimum N is given by $\sigma_{xy}=\frac{e^2}{h}2(N-\frac{1}{2})$,
while for $\text{V}_{\rm{G}}=2$ V, the Hall conductivity is $\sigma_{xy}=\frac{e^2}{h}2N$. }
\end{figure}

\end{document}